\documentclass[preprint,review,12pt]{elsarticle}
\usepackage{amssymb,amsmath,graphicx}
\journal{Physics Letters B}
\usepackage{mathrsfs}
\usepackage{slashed}
\usepackage{bm}
\usepackage{physics}
\usepackage{xcolor}
\usepackage{booktabs}
\usepackage{float}
\usepackage{setspace}
\usepackage{lipsum}
\usepackage{soul}
\usepackage{braket}
\usepackage{tikz}
\usepackage{extarrows}
\usepackage{empheq}
\usepackage{slashed}
\usepackage[
bookmarks=true,
colorlinks,
linkcolor=blue,
urlcolor=blue,
citecolor=blue,
plainpages=false,
pdfpagelabels,
final,
breaklinks=true
]{hyperref}
\usepackage{cleveref}

\crefname{equation}{Eq.}{Eq.}
\crefname{figure}{Fig.}{Fig.}

\begin{document}
	\title{Laser-induced selection of longitudinal \(Z\) bosons in resonant \(e^+e^-\) annihilation}
	\author{Fengye Chen}
	\author{Qingzheng Lv}
	\ead{qzlv@gscaep.ac.cn}
	\author{Libin Fu}
	\ead{lbfu@gscaep.ac.cn}
	\affiliation{Graduate School of China Academy of Engineering Physics, 100193 Beijing, People’s Republic of China}
	\date{\today}
\begin{abstract}

	Longitudinal \(Z\)-boson production is strongly suppressed in collinear \(e^+e^-\) annihilation in vacuum due to current conservation and helicity selection. We show that this suppression is lifted when the incoming leptons are nonperturbatively dressed by an intense circularly polarized plane wave. The dressing produces two distinct effects. The Volkov quasi-momenta shift the collision invariant and allow an initially subresonant collision to reach the \(Z\)-boson pole, whereas the Volkov spinor prefactors generate additional Dirac structures in the neutral current. In the collinear geometry, the field-quadratic contribution to the \(n=0\) harmonic obeys a helicity-selection rule: it vanishes identically for the transverse states (\(\lambda=\pm1\)) of $Z$-boson and survives only for the longitudinal state (\(\lambda=0\)). At sufficiently large \(\xi\) along the physically allowed \(n=0\) resonance trajectory, this helicity-selective contribution becomes dominant, simultaneously driving the growth of the total rate and the enrichment of the longitudinal helicity fraction. We further show that incident-energy asymmetry does not appreciably shift this crossover, because its apparent enhancement of the light-front denominator cancels in the dominant spin-summed matrix element. Instead, the asymmetry controls the laboratory-frame boost of the resonantly produced \(Z\) boson. These results identify a nonlinear helicity-selection mechanism that separates laser control of resonant kinematics from laser control of electroweak-boson polarization.

\end{abstract}

\maketitle

\section{Introduction}
	\label{sec:intro}

	Resonant \(Z\)-boson production is a cornerstone of precision electroweak physics \cite{2006257}. Measurements of its production rate, line shape, total width, and left--right asymmetries have determined the chiral couplings of the neutral current with remarkable accuracy \cite{2006257,abe2000high,RevModPhys.71.575,Acciarri1994,Peter_B_Renton_2002}. Future high-luminosity lepton colliders will extend this program with much larger \(Z\)-boson samples and increasingly precise polarization-sensitive observables \cite{GE2025139269,thecepcstudygroup2025cepctechnicaldesignreport,Benedikt:2928793,Panico:2025kv}. In particular, the longitudinal helicity fraction, accessible through the angular distributions of the \(Z\)-boson decay products, provides a direct probe of contributions that modify the Lorentz and chiral structure of the production current \cite{2023137895,Aaboud2019,Feng2016,Rao2024}.

	Longitudinal \(Z\)-boson production is, however, strongly suppressed in collinear \(e^+e^-\) annihilation in vacuum. For a highly boosted \(Z\) boson, \(\epsilon_0^\mu(K_Z)=K_Z^\mu/m_Z+\mathcal O(m_Z/E_Z)\). Current conservation removes the nominally leading vector-current projection, while the corresponding axial contribution is proportional to the electron mass. Equivalently, the dominant ultrarelativistic initial helicity configurations carry \(J_z=\pm1\) along the collision axis and couple predominantly to the transverse \(Z\)-boson states. This suppression is therefore tied to the kinematics and helicity structure of free incoming leptons. In a coherent background field, both are modified by Volkov dressing, providing a route to alter the longitudinal channel.

\begin{figure}[t!]
		\centering\includegraphics[width=0.98\columnwidth]{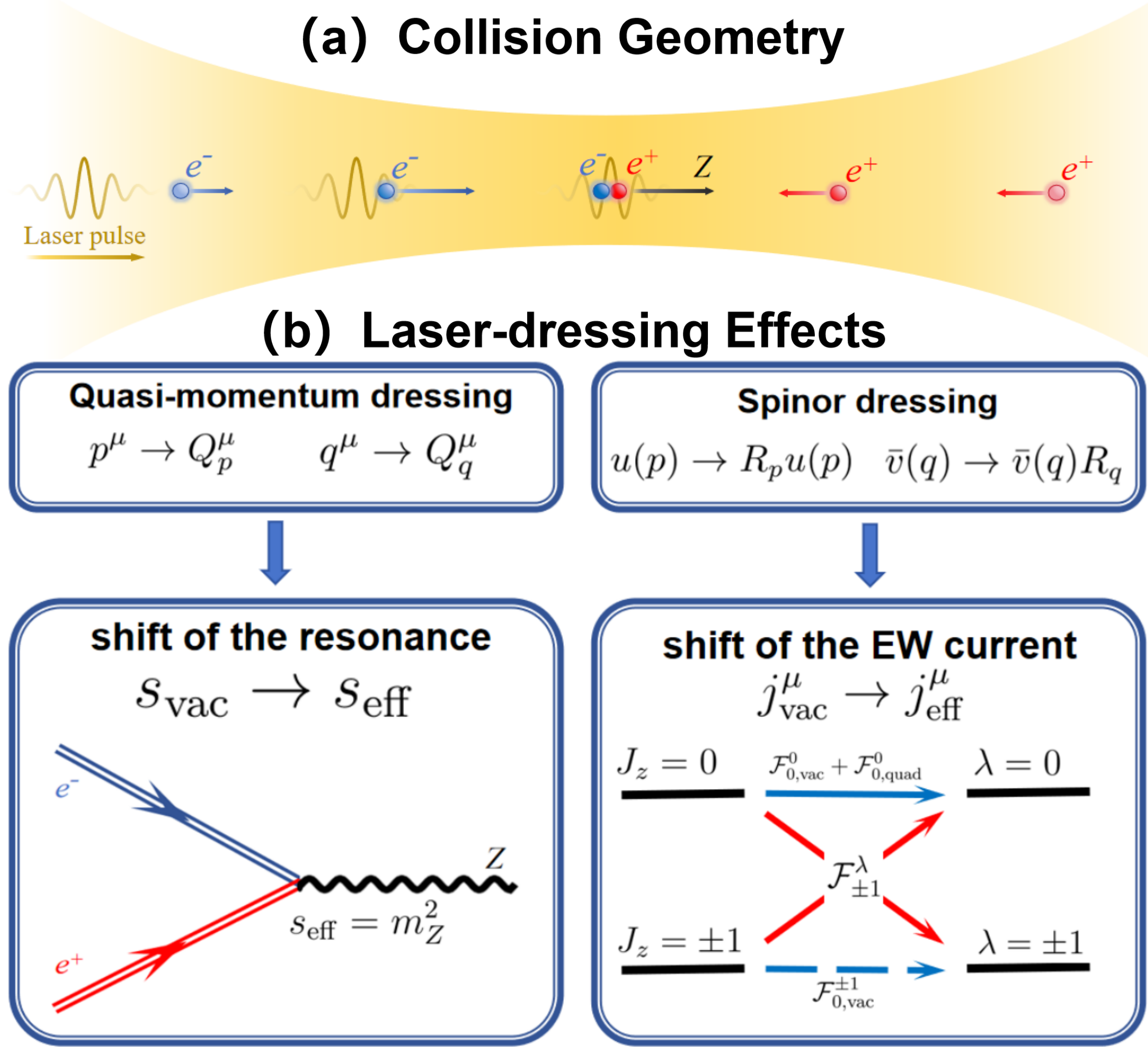}
		\caption{(a) Collision geometry. A circularly polarized laser co-propagates with the incoming electron and counter-propagates against the positron; the dressed pair annihilates inside the field to produce a \(Z\) boson. (b) The two roles of Volkov dressing. Quasi-momentum dressing shifts \(s_{\rm vac}\to s_{\rm eff}\) and enables resonant production at \(s_{\rm eff}=m_Z^2\). Spinor dressing modifies the neutral current and its harmonic helicity amplitudes \(\mathcal F_n^\lambda\). In the collinear geometry, the quadratic \(n=0\) contribution vanishes for \(\lambda=\pm1\) but remains finite for \(\lambda=0\), selectively enhancing the longitudinal channel. Here, \(J_z\) is the initial \(e^+e^-\) angular-momentum projection along the collision axis, and \(\lambda\) denotes the \(Z\)-boson helicity.}
		\label{fig.1}
	\end{figure}

	An intense circularly polarized plane wave modifies the process through two complementary components of Volkov dressing \cite{Wolkow1935,PhysRev.81.115,Ritus1985,RevModPhys.84.1177}. As illustrated in Fig.~\ref{fig.1}, quasi-momentum dressing shifts the effective collision invariant from \(s_{\rm vac}\) to \(s_{\rm eff}\). In the co-/counter-propagating laser beam geometry considered here, the electron co-propagates with the laser, making \(p\cdot k\) small and strongly enhancing its ponderomotive momentum shift. An initially subresonant collision can thereby be brought to \(s_{\rm eff}=m_Z^2\) with $m_Z$ being the $Z$-boson mass. Spinor dressing acts differently: the Volkov prefactors modify the Dirac structure of the neutral current and hence its projections onto the three \(Z\)-boson helicities, without changing the fundamental electroweak couplings.

	Laser-assisted electroweak processes have previously been studied mainly through modifications of production or decay rates \cite{RevModPhys.84.1177,Jakha_2021,Mouslih_2022,Chahri_2023}. The most directly relevant study \cite{OUALI20221182} considered resonant \(e^+e^-\!\to Z\to\mu^+\mu^-\) in a circularly polarized laser field, focusing on the laser-induced modification of the resonant cross section. Related work on laser-assisted Higgs-strahlung, \(e^+e^-\!\to ZH\), likewise examined how the external field modifies the total production cross section \cite{Ouhammou_2021}. The polarization composition of the produced \(Z\), and in particular the redistribution between longitudinal and transverse channels induced by Volkov dressing, was not addressed in these studies.

	Here we demonstrate that the two effects control distinct observables. The quasi-momentum shift determines access to the \(Z\)-boson pole, whereas the polarization redistribution originates from the nonlinear dressed current. In the collinear geometry, the field-quadratic \(n=0\) contribution vanishes identically for both transverse helicity $Z$-boson but remains finite for the longitudinal state. Once this contribution becomes important, it drives both the high-field growth of the total rate and the enrichment of the longitudinal channel. Incident-energy asymmetry does not appreciably lower the corresponding field scale, but instead controls the laboratory-frame boost of the on-shell \(Z\) boson. The field strength and collision asymmetry therefore provide distinct controls over the polarization composition and final-state motion.

\section{Laser-dressed \(Z\)-boson production}
	\label{sec:theory}

	We formulate laser-assisted resonant \(e^{+}e^{-}\) annihilation within the Furry picture \cite{RevModPhys.84.1177,PhysRev.81.115}, treating the coupling of the incoming leptons to the circularly polarized laser field nonperturbatively while retaining the electroweak interaction to leading order. The relevant Lagrangian is
	\begin{equation}
		\begin{split}
			\mathcal L
			=&\bar\psi\left(i\slashed{\partial}	- m_e - e_L\slashed A\right)\psi - \frac{1}{4}\mathcal Z^{\mu\nu}\mathcal Z_{\mu\nu} + \frac{1}{2}m_Z^2 Z^\mu Z_\mu
			\\
			&- \frac{g}{2\cos\theta_W} \bar\psi \gamma^\mu \left(g_V-g_A\gamma^5\right)\psi Z_\mu ,
		\end{split}
		\label{eq:lagrangian-furry}
	\end{equation}
	where \(\mathcal Z^{\mu\nu}=\partial^\mu Z^\nu-\partial^\nu Z^\mu\). We use an effective-Born \(Z\)-pole input scheme with \(\alpha^{-1}(m_Z^2)=128.95\) and \(\sin^2\theta_W\equiv\sin^2\theta_{\mathrm{eff}}^\ell=0.23154\) \cite{2006257}. Accordingly, \(e_Z=-\sqrt{4\pi\alpha(m_Z^2)}\), \(g=|e_Z|/\sin\theta_W\), \(g_V=-1/2+2\sin^2\theta_W\), and \(g_A=-1/2\). The optical background instead couples through \(e_{\rm L}=-\sqrt{4\pi\alpha(0)}\), with \(\alpha^{-1}(0)=137.036\); thus, \(e_{\rm L}\neq e_Z\) accounts for the running of the electromagnetic coupling. No electroweak corrections beyond this effective-Born prescription are included, and all laser dependence enters through the Volkov-dressed external states. We define \(a\cdot b=g_{\mu\nu}a^\mu b^\nu\) and use \(g_{\mu\nu}=\operatorname{diag}(+,-,-,-)\) and natural units throughout.

	The circularly polarized background is modeled as a monochromatic plane wave, \(A^\mu(\phi)=a\bigl(\epsilon_{\rm L}^\mu e^{i\phi}+\epsilon_{\rm L}^{\mu}e^{-i\phi}\bigr)\), with \(\phi=k\cdot x\), \(k^\mu=\omega(1,0,0,1)\), and \(k\cdot\epsilon_{\rm L}=0\), \(\epsilon_{\rm L}^2=0\), \(\epsilon_{\rm L}\cdot\epsilon_{\rm L}^=-1/2\). The field strength is characterized by \(\xi=|e_{\rm L}|a/m_e\) with $m_e$ being the electron mass. We neglect variations of the laser envelope and transverse profile across the microscopic formation region of the hard annihilation process \cite{Ritus1985,RevModPhys.84.1177,RevModPhys.94.045001}.

	The incoming electron is described by the Volkov state in a plane wave as $\psi^{(-)}_{p,s}(x)=R_p(\phi)u_s(p)e^{-iS_p(x)}/\sqrt{2Q_p^0V}$ with $R_p(\phi)=1+e_{\rm L}\slashed{k}\slashed{A}(\phi)/(2p\cdot k)$. The positron state is obtained analogously by charge conjugation. The secular terms of the Volkov phases define the cycle-averaged quasi-momenta $Q_\ell^\mu=\ell^\mu+\Delta^\mu(\ell)=\ell^\mu+m_e^2\xi^2/(2\ell\cdot k)\,k^\mu$ with $\ell=p,q$ for electron and positron, respectively. They both satisfy \(Q_p^2=Q_q^2=m_*^2=m_e^2(1+\xi^2)\). These quasi-momenta are the Volkov-dressed effective momenta entering energy--momentum conservation at the annihilation vertex, rather than asymptotic momentum shifts of the incoming leptons.

	For a channel involving a net exchange of \(n\) laser photons, the effective invariant energy is
	\begin{equation}
		\begin{split}
			s_{\rm eff, n}&=\left(Q_p+Q_q+nk\right)^2\\
			&=s_{\mathrm{vac}}+m_e^2\xi^2\left(2+\frac{p\cdot k}{q\cdot k}+\frac{q\cdot k}{p\cdot k}\right)+2n\,k\cdot(p+q),
		\end{split}
		\label{eq:effective-s-channel}
	\end{equation}
	where \(s_{\rm vac}=(p+q)^2\). In the co-/counter-propagating geometry, the large ratio \(q\cdot k/(p\cdot k)\) strongly enhances the quasi-momentum contribution, allowing an initially subresonant collision to be tuned to the \(Z\)-boson pole, \(s_{\rm eff,0}\simeq m_Z^2\) \cite{supplementary}.

	The second effect of Volkov dressing enters through the spinor prefactors, which modify the leptonic neutral current according to
	\begin{equation}
		J_{\mathrm{eff}}^\mu(\phi)=\bar v(q)\bar R_q(\phi)\gamma^\mu\left(g_V-g_A\gamma^5\right)R_p(\phi)u(p),
		\label{eq:J_eff}
	\end{equation}
	with \(\bar R_q=\gamma^0R_q^\dagger\gamma^0\) and $R_q(\phi)$ for positron states. Separating the quasi-momentum phases and Fourier expanding the remaining \(\phi\)-periodic factors then gives the Furry-picture \(S\)-matrix
	\begin{equation}
		\begin{split}
		S_{fi}^{\lambda}&=(2\pi)^4 \sum_n\delta^{(4)}\left(Q_p+Q_q+nk-K_Z\right)i\mathcal{M}_{n,s's}^{\lambda} \\
		&=-i\frac{(2\pi)^4g}{2\cos\theta_W} \sum_n\delta^{(4)}\left(Q_p+Q_q+nk-K_Z\right)\bar{v}_{s'}(q)\,\mathcal{F}_n^\lambda\,u_s(p).
		\end{split}
		\label{eq:S-matrix-channel}
	\end{equation}
	where $\mathcal F_n^\lambda=\left[\bar R_q(\phi)\slashed{\epsilon}_{\lambda}^{*}(K_Z)(g_V-g_A\gamma^5)R_p(\phi)e^{i\Phi(\phi)}\right]_n$, \(K_Z^2=m_Z^2\) and \(\lambda=0,\pm1\) labels the \(Z\)-boson helicity. Here \(\Phi(\phi)\) denotes the periodic part of the combined Volkov phase and \([\cdots]_n\) its \(n\)th Fourier coefficient.

	In a general collision geometry, the Fourier coefficients contain Bessel functions describing multiphoton exchange with the background \cite{supplementary}. In the present collinear geometry, however, the multiphoton parameter \(\zeta=2m_e\xi\left|p\cdot\epsilon_{\rm L}/(p\cdot k)-q\cdot\epsilon_{\rm L}/(q\cdot k)\right|=0\) because \(p\cdot\epsilon_{\rm L}=q\cdot\epsilon_{\rm L}=0\), so the harmonic structure reduces to
	\begin{equation}
		\begin{aligned}
			\mathcal F_{-1}^{\lambda}
			&=
			-m_e\xi
			\left[
			\frac{G^\lambda\slashed{k}\slashed{\epsilon}_{\rm L}^{*}}
				{2p\cdot k}
			-
			\frac{\slashed{\epsilon}_{\rm L}^{*}\slashed{k}G^\lambda}
				{2q\cdot k}
			\right],
			\\[1mm]
			\mathcal F_{0}^{\lambda}
			&=
			G^\lambda
			+
			\frac{m_e^2\xi^2}{4(p\cdot k)(q\cdot k)}
			\Big[
			\slashed{k}\slashed{\epsilon}_{\rm L}
			G^\lambda
			\slashed{k}\slashed{\epsilon}_{\rm L}^{*}
			+
			\slashed{k}\slashed{\epsilon}_{\rm L}^{*}
			G^\lambda
			\slashed{k}\slashed{\epsilon}_{\rm L}
			\Big],
			\\[1mm]
			\mathcal F_{+1}^{\lambda}
			&=
			-m_e\xi
			\left[
			\frac{G^\lambda\slashed{k}\slashed{\epsilon}_{\rm L}}
				{2p\cdot k}
			-
			\frac{\slashed{\epsilon}_{\rm L}\slashed{k}G^\lambda}
				{2q\cdot k}
			\right].
		\end{aligned}
		\label{eq:collinear-harmonics}
	\end{equation}
	where \(G^\lambda=\slashed{\epsilon}_{\lambda}^{*}(K_Z)(g_V-g_A\gamma^5)\). Thus, the \(n=\pm1\) kernels are linear in \(\xi\), whereas the \(n=0\) channel contains both the vacuum term and a quadratic Volkov contribution. Using \(k^2=k\cdot\epsilon_{\rm L}=0\) and \(\slashed{\epsilon}_{\rm L}\slashed{\epsilon}_{\rm L}^{*}+\slashed{\epsilon}_{\rm L}^{*}\slashed{\epsilon}_{\rm L}=-1\), the latter becomes $\mathcal F_{0,\mathrm{quad}}^\lambda=-m_e^2\xi^2\left(k\cdot\epsilon_\lambda^{*}\right)\slashed{k}(g_V-g_A\gamma^5)/[2(p\cdot k)(q\cdot k)]$. For collinear \(K_Z\) and \(k\), \(k\cdot\epsilon_{\pm1}=0\) while \(k\cdot\epsilon_0\neq0\), and therefore
	\begin{equation}
		\mathcal F_{0,\mathrm{quad}}^{\pm1}=0,
		\qquad
		\mathcal F_{0,\mathrm{quad}}^{0}\neq0.
		\label{eq:longitudinal-selection-rule}
	\end{equation}
	The quadratic \(n=0\) contribution therefore obeys a strict helicity selection rule: it contributes only to the longitudinal channel. The resulting longitudinal enrichment is thus a dynamical consequence of the nonlinear Volkov current rather than a kinematic consequence of tuning \(s_{\rm eff,0}\) to the \(Z\) pole.

	Because the unstable \(Z\) boson is not an asymptotic state, we define the observable through the inclusive resonant process \(e^+e^-\to Z \to X\). In the pole approximation, the helicity-resolved production residue is evaluated at \(K_Z^2=m_Z^2\), while the finite lifetime is retained through the Breit--Wigner denominator \cite{PhysRev.49.519,Maggiore2004}, giving
	\begin{equation}
		\sigma^{\lambda} = \sum_n \frac{1}{2\mathcal I_{\rm in}}\, \overline{\sum_{s,s'}} \left| \mathcal M_{n,s's}^{\lambda} \right|^2 \frac{m_Z\Gamma_Z} {(s_{\rm eff, n}-m_Z^2)^2+m_Z^2\Gamma_Z^2}.
		\label{eq:polarized-cross-section}
	\end{equation}
	Here \(\mathcal I_{\rm in}=\sqrt{(Q_p\cdot Q_q)^2-m_*^4}\) is the invariant flux and \(\overline{\sum}_{s,s'}=\frac14\sum_{s,s'}\) denotes the initial-spin average. The normalization is chosen consistently with the Volkov-state normalization so that the helicity-summed result recovers the inclusive Born-level \(Z\)-pole cross section as \(\xi\to0\). We use the vacuum width \(\Gamma_Z\), since background-field corrections to the pole width remain negligible in the collinear kinematics considered here \cite{supplementary,Jakha_2021}. Finally,
	\begin{equation}
		\begin{aligned}
			\sigma_{\rm tot} &= \sum_{\lambda=-1}^{1} \sigma^{\lambda}, \qquad
			R_\lambda &= \frac{\sigma^{\lambda}}{\sigma_{\rm tot}}.
		\end{aligned}
		\label{eq:total-cross-section_and_polarization-fraction}
	\end{equation}
	Thus, the intensity dependence of \(\sigma_{\rm tot}\) reflects both resonance tuning and dressing of the production amplitudes, whereas \(R_\lambda\) isolates the corresponding helicity redistribution.

\section{Cross section and longitudinal-polarization enhancement}
	\label{sec:results}

	\begin{figure}[t!]
		\centering\includegraphics[width=0.98\columnwidth]{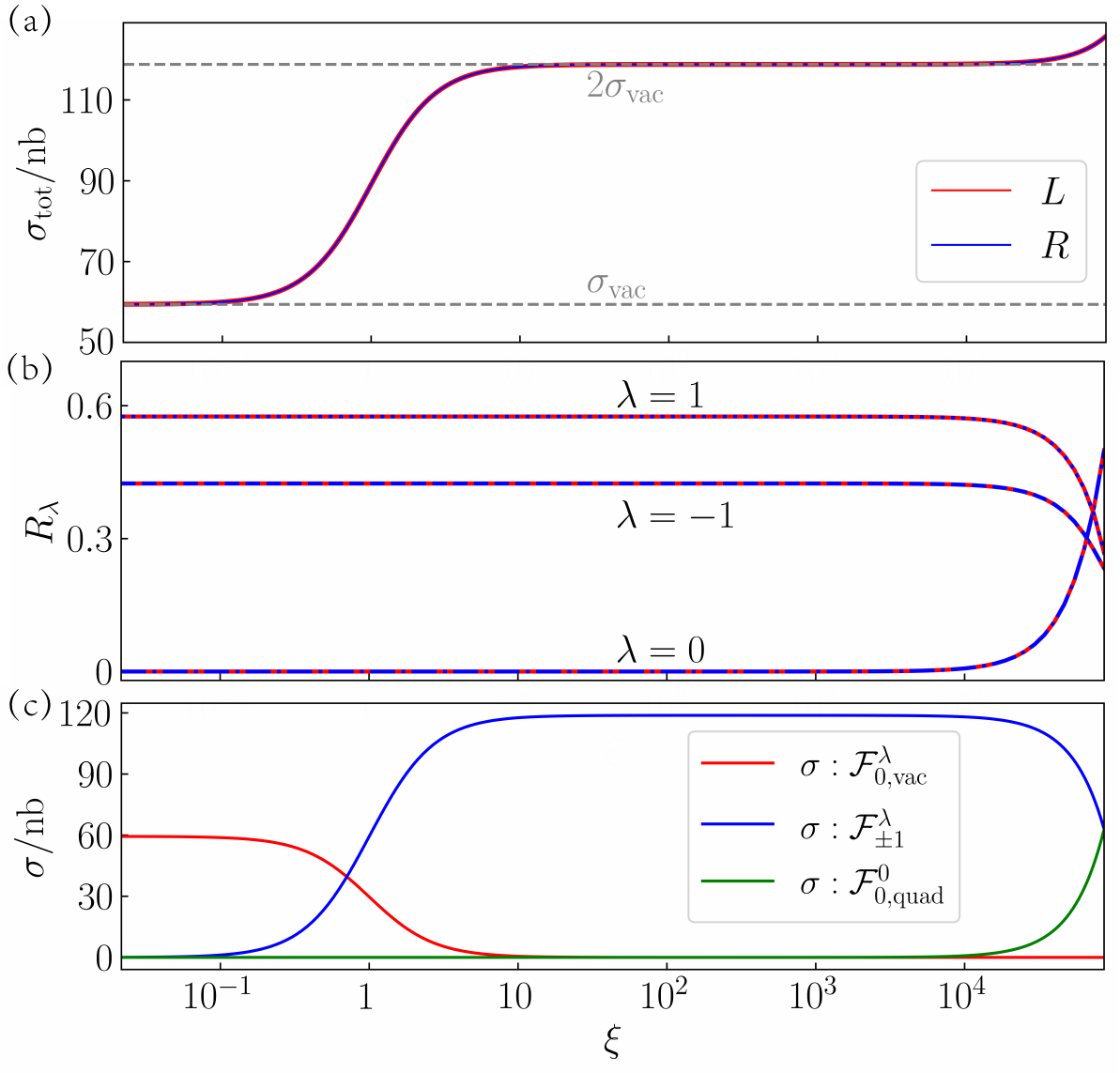}
		\caption{(a) Total cross section \(\sigma\) as a function of the laser intensity parameter \(\xi\) for left- (\(L\), red) and right-handed (\(R\), blue) circularly polarized laser fields. The dashed lines indicate the vacuum cross section \(\sigma_{\mathrm{vac}}\) and \(2\sigma_{\mathrm{vac}}\). (b) Corresponding helicity fractions \(R_\lambda\) for \(\lambda=0,\pm1\). The longitudinal fraction \(R_0\), negligible at low and intermediate \(\xi\), increases strongly at high field strengths, accompanied by a reduction of the transverse fractions. The results for the two laser helicities nearly coincide in (a) and (b). (c) Diagnostic contributions from the vacuum \(n=0\), linear \(n=\pm1\), and quadratic \(n=0\) amplitude structures, which predominantly govern the vacuum limit, intermediate-field plateau, and high-field upturn, respectively. The vacuum and quadratic terms interfere coherently within the \(n=0\) channel, but the interference contribution is numerically small over the range shown and is not displayed separately.}
		\label{fig.2}
	\end{figure}

	We isolate the modification of the production current by following the \(n=0\) resonance trajectory, \(s_{\rm eff,0}=m_Z^2\), adjusting the equal incident energies at each \(\xi\) so that the dressed collision remains on the \(Z\) pole. The scan therefore does not describe an approach to resonance, but compares the dressed helicity amplitudes under a common resonant constraint. For symmetric incidence [$p^0=q^0=\frac{m_Z}{2\sqrt{1+\xi^2}}$], the propagating incident beams require $\xi<\xi_{\max}=\sqrt{m_Z^2/(4m_e^2)-1}\simeq8.9\times10^4$. The endpoint \(\xi=\xi_{\max}\) corresponds to vanishing incident three-momenta and hence marks the kinematic boundary of the co-/counter-propagating beam geometry; Fig.~\ref{fig.2} shows only the physical results below this limit.

	The evolution of the total rate reflects the successive dominance of the dressed-current structures in Eq.~\eqref{eq:collinear-harmonics}. For \(\xi\lesssim10^{-1}\), \(\sigma_{\rm tot}\) approaches the field-free value \(\sigma_{\rm vac}=59.6~\mathrm{nb}\) \cite{Thomson:2013}. Over \(10\lesssim\xi\lesssim10^4\), the field-linear \(n=\pm1\) channels dominate and produce the broad plateau at approximately \(2\sigma_{\rm vac}\); its weak \(\xi\)-dependence results from compensating changes of the linear kernels, incident energies, state normalization, phase space, and invariant flux along the resonance trajectory \cite{supplementary}. At larger \(\xi\), the quadratic \(n=0\) amplitude becomes comparable to and eventually dominates the dressed response, producing the high-field upturn with an approximately quadratic dependence on \(\xi\). Since \(\mathcal F_{0,\mathrm{vac}}^\lambda\) and \(\mathcal F_{0,\mathrm{quad}}^\lambda\) belong to the same harmonic channel, the full \(n=0\) contribution contains their coherent interference, although it is numerically small [Fig.~\ref{fig.2}(c)]. Reversing the laser helicity exchanges the \(n=\pm1\) kernels and hence the two transverse-helicity labels, while leaving the helicity-summed rate and \(R_0\) invariant, providing a symmetry check of the calculation [Figs.~\ref{fig.2}(a) and \ref{fig.2}(b)].

	The helicity fractions expose the qualitative effect of the quadratic current that is not apparent from the total rate alone. Across the intermediate-field plateau, the linear \(n=\pm1\) channels cause only a modest helicity redistribution and \(R_0\) remains close to its field-free value. Once the quadratic \(n=0\) contribution becomes important for \(\xi\gtrsim10^4\), however, the selection rule in Eq.~\eqref{eq:longitudinal-selection-rule} enhances only the longitudinal channel. Consequently, \(R_0\) rises rapidly toward \(0.50\) near the upper end of the physical resonance domain, while the transverse fractions decrease. The high-field growth of the total rate and the longitudinal enrichment are therefore two manifestations of the same nonlinear Volkov-current contribution.

	\begin{figure}[t!]
		\centering\includegraphics[width=0.98\columnwidth]{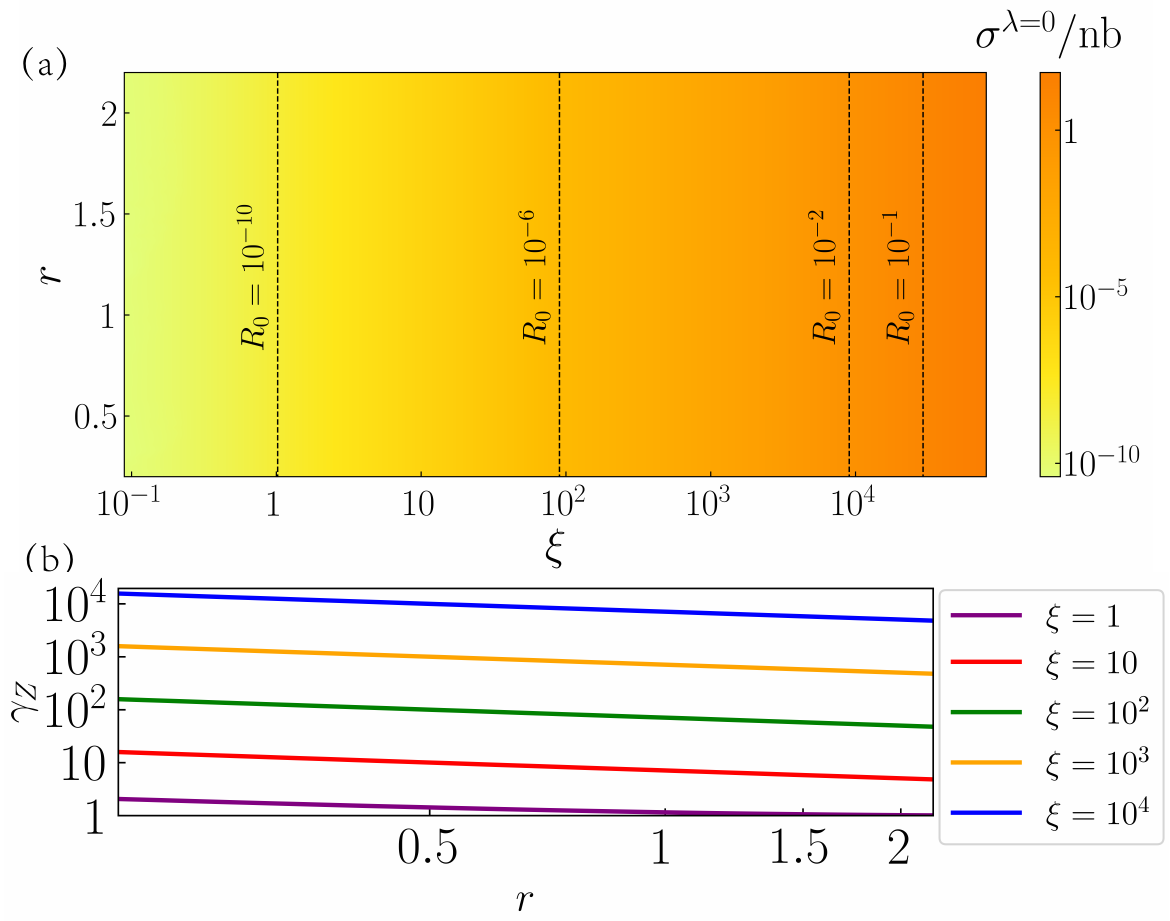}
		\caption{(a) Longitudinal \(Z\)-boson production cross section \(\sigma_{\lambda=0}\) as a function of the laser intensity parameter \(\xi\) and the incident-energy ratio \(r=\abs{\vec q}/\abs{\vec p}\), evaluated under the dressed resonance condition \(s_{\mathrm{eff,0}}=m_Z^2\). The color scale and labeled contours give \(\sigma_{\lambda=0}\) in nb. The nearly vertical contours for the longitudinal fraction $R_0$ show that varying the incident-energy asymmetry does not appreciably change the field strength required for longitudinal enhancement. (b) Lorentz factor \(\gamma_Z\) of the produced on-shell \(Z\) boson as a function of \(r\) for \(\xi=10\), \(10^2\), \(10^3\), and \(10^4\). Decreasing \(r\) produces increasingly boosted \(Z\) bosons, with the boost increasing further with \(\xi\). Thus, incident-energy asymmetry controls the final-state boost while preserving the enhancement of longitudinal \(Z\)-boson production.}
		\label{fig.3}
	\end{figure}

	The quadratic \(n=0\) kernel contains the light-front factor \(1/[(p\cdot k)(q\cdot k)]\) [Eq.~\eqref{eq:collinear-harmonics}], suggesting that incident-energy asymmetry might enhance the longitudinal contribution. Defining \(r\equiv|\bm q|/|\bm p|\simeq q^0/p^0\), the relativistic collinear limit gives \(p\cdot k\simeq\omega m_e^2/(2p^0)\) and \(q\cdot k\simeq2\omega q^0\), so that \(1/[(p\cdot k)(q\cdot k)]\simeq1/(\omega^2m_e^2r)\). At the kernel level, decreasing \(r\) therefore appears to produce a \(1/r\) enhancement.

	This kernel-level scaling does not carry over to the observable rate because the incident energies are constrained by the dressed resonance condition. In the relativistic collinear limit, \(s_{\rm eff,0}=m_Z^2\), one has \(p^0q^0\simeq m_Z^2/[4(1+\xi^2)]\), so varying \(r\) redistributes a fixed resonant energy product rather than moving the collision closer to the pole. In light-front coordinates \(v^\pm=v^0\pm v^3\), with \(k\cdot v=\omega v^-\), the resonance condition gives \((p^-+q^-)^2/(p^-q^-)=m_Z^2/m_*^2\). Since \(k\cdot\epsilon_0=-\omega(p^-+q^-)/m_Z\) and \((p\cdot k)(q\cdot k)=\omega^2p^-q^-\), it follows that \((k\cdot\epsilon_0)^2/[(p\cdot k)(q\cdot k)]=1/m_*^2\). Combining this identity with \(\frac14\sum_{\rm spins}|\bar v(q)\slashed{k}(g_V-g_A\gamma^5)u(p)|^2 =2(g_V^2+g_A^2)(p\cdot k)(q\cdot k)\) yields
	\begin{equation}
		\frac14\sum_{\rm spins}\left|\bar v(q)\mathcal F_{0,\mathrm{quad}}^{0}u(p)\right|^2=\frac{g_V^2+g_A^2}{2}\frac{m_e^2\xi^4}{1+\xi^2},
		\label{eq:r-cancellation}
	\end{equation}
	apart from the overall neutral-current coupling. The spin-summed current therefore cancels the apparent \(1/r\) enhancement, leaving the dominant quadratic longitudinal contribution controlled primarily by \(\xi\). This explains the nearly vertical \(\sigma^{\lambda=0}\) bands and fixed-\(R_0\) contours in Fig.~\ref{fig.3}(a); the weak residual \(r\)-dependence arises from the subleading vacuum, field-linear, and interference terms.

	The asymmetry nevertheless controls the laboratory-frame boost of the produced \(Z\) boson. Using \(q^0\simeq rp^0\) together with \(p^0q^0\simeq m_Z^2/[4(1+\xi^2)]\) gives
	\begin{equation}
		\gamma_Z=\frac{K_Z^0}{m_Z}\simeq\frac{1}{2}\left[\sqrt{\frac{1+\xi^2}{r}}+\sqrt{\frac{r}{1+\xi^2}}\right]\simeq\frac{1}{2}\sqrt{\frac{1+\xi^2}{r}},
		\label{eq:Z-boost}
	\end{equation}
	where the final form applies for \((1+\xi^2)/r\gg1\) while both beams remain relativistic; outside this regime, the physical boundaries must be obtained from the exact resonance kinematics. Figure~\ref{fig.3}(b) therefore shows that decreasing \(r\) can substantially increase the \(Z\)-boson boost without appreciably shifting the onset of longitudinal enhancement. Thus, \(\xi\) controls the helicity-selective quadratic current, whereas \(r\) primarily controls the laboratory-frame motion of the produced \(Z\) boson.

\section{Discussion and conclusions}
	\label{sec:discussion_conclusion}

	The preceding results were obtained for monochromatic, perfectly collinear incident beams. Because the longitudinal enhancement relies on both resonant tuning and a collinear helicity-selection rule, we have estimated the leading sensitivity to finite angular divergence and energy spread by combining beam-distribution averaging with analytic bounds on harmonic detuning and transverse-helicity leakage \cite{supplementary}. Angular divergence activates neighboring harmonics over an effective range \(|n|\lesssim\zeta\), but this broadening primarily redistributes the harmonic strength; the variation of the associated Breit--Wigner weights is controlled by \(\eta_\Gamma=2\zeta|k\cdot K_{Z,0}|/(m_Z\Gamma_Z)\). Finite transverse momentum weakly relaxes the collinear selection rule, with	\(\sigma_{0}^{\pm1}(\mathcal F_{0,\mathrm{quad}}^{\pm1})/\sigma_{0}^{0}(\mathcal F_{0,\mathrm{quad}}^{0}) =O(K_{Z,\perp}^2/m_Z^2)\). Along the symmetric resonance trajectory, root mean square (rms) divergences \(\sigma_{\theta}\lesssim1~\mathrm{mrad}\) give \(|K_{Z,\perp}|\lesssim O(10~\mathrm{keV})\), \(\eta_\Gamma\lesssim10^{-4}\), and transverse leakage at the \(10^{-14}\) level. Energy spread does not open the forbidden transverse quadratic channel, although it broadens \(s_{\rm eff,0}\) across the resonance. Requiring the Breit--Wigner weight at one rms displacement to remain within \(1\%\) (\(10\%\)) of its peak gives a corresponding single-beam fractional-spread estimate of \(0.27\%\) (\(0.91\%\)). A systematic shift of the mean energies can be compensated by retuning the laser intensity, whereas stochastic broadening primarily reduces the resonantly weighted yield. Once the quadratic \(n=0\) term dominates throughout the resonant interval, the longitudinal fraction is expected to be less sensitive than the total resonantly weighted rate.

	Taken together, our results show that an intense circularly polarized optical field controls resonant \(Z\)-boson production through two distinct aspects of Volkov dressing. The laser-induced quasi-momenta shift the effective invariant energy and can tune an initially subresonant \(e^+e^-\) collision onto the \(Z\)-boson pole. The Volkov spinor prefactors instead generate additional Dirac structures in the dressed neutral current and redistribute the \(Z\)-boson helicities. The dynamical origin of this redistribution is exposed by the quadratic \(n=0\) contribution, which obeys the strict collinear selection rule \(\mathcal F_{0,\mathrm{quad}}^{\pm1}=0\), while \(\mathcal F_{0,\mathrm{quad}}^{0}\neq0\). Once dominant, this contribution drives both the high-field growth of the total rate and the enrichment of the longitudinal channel. The latter is thus not a kinematic by-product of resonance tuning, but a helicity-selection effect generated by the nonlinear Volkov current.

	Incident-energy asymmetry provides complementary control over the laboratory-frame motion of the produced boson. Although the quadratic kernel contains an explicit light-front denominator that appears to scale as \(1/r\), this dependence cancels in the dominant spin-summed longitudinal matrix element along the resonance trajectory. Varying \(r\) therefore does not appreciably lower the field scale at which the helicity-selective contribution becomes important, but can substantially increase the boost of the on-shell \(Z\) boson. Along the resonance trajectory, \(\xi\) and \(r\) thus govern primarily the polarization composition and laboratory-frame boost, respectively, identifying a route toward strongly boosted, longitudinally enriched \(Z\) bosons. The present calculation establishes this separation at leading order in the Furry picture; translating the idealized cross sections into realistic event yields will additionally require finite-pulse effects, beam luminosity, and radiative survival probabilities.

	This work was supported by the National Natural Science Foundation of China (Grants No. 12088101 and No. U2330401).
	\bibliographystyle{unsrt}
	\bibliography{bib.bib}
\end{document}